\documentclass[a4paper,11pt]{article}
\usepackage{pos}
\usepackage{psfrag}
\usepackage{multirow}

\title{Accessing Generalized Parton Distributions through $2 \to 3$ exclusive processes}

\author[1]{Goran Duplan\v{c}i\'{c}}
\author[2,3]{Saad Nabeebaccus}
\author[1]{Kornelija Passek-K.}
\author[4]{Bernard Pire}
\author[5]{Lech Szymanowski}
\author*[2]{Samuel Wallon}

\affiliation[1]{Theoretical Physics Division, Rudjer Bo{\v s}kovi{\'c} Institute,
	HR-10002 Zagreb, Croatia}
\affiliation[2]{Universit\'e Paris-Saclay, CNRS/IN2P3, IJCLab, 91405 Orsay, France}
\affiliation[3]{Department of Physics \& Astronomy, University of Manchester, Manchester M13 9PL, United Kingdom}
\affiliation[4]{CPHT, CNRS, Ecole polytechnique, Institut Polytechnique de Paris, 91128 Palaiseau, France}
\affiliation[5]{National Centre for Nuclear Research (NCBJ), 02-093 Warsaw, Poland}

\abstract{We review our results on a new class of $2 \to 3$ exclusive processes, as a probe of both chiral-even and chiral-odd quark GPDs. We consider the exclusive photoproduction of a photon-meson pair, in the kinematics where the pair has a large invariant mass, described in the collinear factorization framework. We cover the whole kinematical range from medium energies in fixed target experiments to very large energies of colliders, by considering the experimental conditions of JLab 12-GeV, COMPASS, future EIC and LHC (in ultra-peripheral collisions) cases. Our analysis covers neutral and charged rho-mesons, as well as charged pions. The case of the rho-meson, depending on its polarization, provides access to either chiral-even or chiral-odd GPDs, at leading twist. We find that the order of magnitude of the obtained cross sections are sufficiently large for a dedicated experimental analysis to be performed, especially at JLab. Furthermore, we compute the linear photon beam polarization asymmetry, which we find to be sizeable, in the case of a longitudinally polarized $\rho$-meson or of a charged pion. These predictions are obtained for both asymptotic distribution amplitude (DA) and holographic DA.}

\FullConference{31st International Workshop on Deep Inelastic Scattering (DIS2024)\\
8–12 April 2024\\
Grenoble, France\\}

\begin{document}
\maketitle

\section{Introduction}

Two decades ago, exclusive processes opened the door to a possible access to three-dimensional tomography of the internal content of nucleons in terms of partons. A key tool, for hard processes, is the use of factorization theorems which allows the description of the scattering amplitude of these processes through a convolution of a short distance coefficient function and Generalised Parton Distributions (GPDs). GPDs are defined through non-forward matrix elements of quark-antiquark operators or gluon field strength operators, separated by a light-like distance. At the leading power in the hard scale (twist-2), there are 8 quark GPDs and 8 gluonic GPDs~\cite{Diehl:2003ny}. In the quark sector, they split into 4 chiral-even ones, denoted by $ H_{q},\, E_{q},\,\tilde{H}_{q}  $ and $  \tilde{E}_{q}  $, and 4 chiral-odd (helicity flip) ones, denoted by $ H_{q}^{T},\, \tilde{H}^{T}_{q},\,E_{q}^{T}  $ and $  \tilde{E}_{q}^{T} $.  A GPD depends on 3 variables: $ x $ (the average fraction of the nucleon longitudinal momentum carried by the probed quark/antiquark), $  \xi  $, the so-called skewness parameter (twice the difference in longitudinal momentum fraction between the probed partons), and $ t $ (momentum transfer squared). At $  \xi =0$, 
the Fourier transform (wrt $ t $) of GPDs gives the probability distribution of partons in the transverse plane \cite{Burkardt:2000za,Burkardt:2002hr}.

At the moment, in order to extract GPDs, the focus has been mainly on 2 $\to$ 2
scattering processes, such as deeply-virtual Compton scattering (DVCS) and deeply-virtual meson
production (DVMP), where collinear factorisation has explicitly been shown to hold at all order in
perturbation theory (in $\alpha_s$) at the leading twist. In DVMP, the amplitude factorises, for specific polarisations of the incoming photon and outgoing meson, in terms of a coefficient function, a GPD, and a distribution amplitude (DA) for the outgoing meson.

To probe chiral-odd GPDs, since in the massless limit QED and QCD are chiral-even theories, another source of odd chirality should appear in the amplitude, in order to get a non-zero result at leading twist. Unfortunately, a simple 2 $\to$ 2 process like the exclusive electroproduction of a transversely polarised $  \rho_{T}  $-meson (the DA of the $  \rho  $-meson at the leading twist-2 is chiral-odd, thus promising) vanishes at all orders in $  \alpha _{s} $ since it would require a helicity transfer of 2 units from a photon \cite{Diehl:1998pd,Collins:1999un}. To evade this, 
it was proposed in \cite{Ivanov:2002jj,Enberg:2006he,
	ElBeiyad:2010pji} to consider the photo- and electro-production of two mesons, through a 3-body final state. Similarly, the study of the photoproduction of a $ \gamma  \rho _{T} $ pair was proposed~\cite{Boussarie:2016qop}, the large (and almost opposite) transverse momenta of the photon and meson in the final state providing the hard scale for collinear factorisation of the process \cite{Qiu:2022bpq,Qiu:2022pla}. Diphoton photoproduction was also studied to access chiral-even GPDs~\cite{Grocholski:2022rqj}. We here report on our analyses
	of the photoproduction of a photon-meson pair, the meson in the final state being either a charged pion, or a $\rho$-meson of any charge and polarisation, at leading order (LO)~\cite{Duplancic:2018bum,Duplancic:2022ffo,Duplancic:2023kwe}. 
	Note that these channels, because of involved quantum numbers, do not  allow for two-gluon exchange in $t-$channel. 
	
	Besides the possibility of accessing chiral-odd GPDs, 3-body final states provide a natural hard scale, even in photoproduction, a particularly appealing fact in ultraperipheral collisions (UPC) at the LHC. Furthermore, they give access to an enhanced $ x $-dependence of GPDs, in comparison with simple exclusive $2 \to 2$ processes like DVCS which only effectively probe the GPD at $ x = \pm \xi $ at LO~\cite{Qiu:2023mrm}. Note that double deeply-virtual Compton scattering, due to the presence of two scales (initial and final photon virtualities) also allows for such a scan of the $x$-dependence of GPDs~\cite{Deja:2023ahc}.
	
	A proof of factorisation was given for such a class of $ 2\to 3 $ exclusive processes~\cite{Qiu:2022bpq,Qiu:2022pla}. Still, for the exclusive photoproduction of a $\pi ^{0}\gamma  $ pair, which allows for 
	a gluon GPD contribution,  
	a pure collinear factorisation approach is invalid, caused by the existence of a \textit{Glauber pinch}~\cite{Nabeebaccus:2023rzr,Nabeebaccus:2024mia,Nabeebaccus:2024rig}.

\section{Photoproduction of $  \gamma \pi ^{\pm} $ and $ \gamma  \rho _{L,T}^{0,\,\pm} $ with large invariant mass}

\label{sec:photon-meson-photoproduction}

\subsection{Kinematics}

We consider the process
\begin{align}
	\label{eq:process}
	\gamma (q) + N(p_N) \to \gamma (k) + M (p_{M}) + N' (p_{N'})\,.
\end{align}
Any generic momentum $ r $ can be written as $r^{ \mu } = a p^{  \mu } + b n^{ \mu } + r^{ \mu }_{\perp}\,,$ using the
lightcone vectors $p^{ \mu } = \frac{\sqrt{s}}{2} \left( 1,0,0,1 \right)$ and
$n^{ \mu } = \frac{\sqrt{s}}{2} \left( 1,0,0,-1 \right)$
with $ n  \cdot p = \frac{s}{2} $,
with the convention $ r_{\perp}^2 \equiv - |\vec{r}_{t}|^2 $. Then,
\begin{align}
p_{N} &=  \left( 1+ \xi \right) p^{ \mu } + \frac{m_{N}^2}{s \left( 1+\xi \right) }n^{ \mu }\,,\quad\;\;
p_{N'}=\left( 1- \xi \right) p^{ \mu } + \frac{m_{N}^2+ |\vec{\Delta} _{t}|^2}{s \left( 1-\xi \right) }n^{ \mu }+ \Delta _{\perp}^{ \mu }\,,\quad\;\;
q^{ \mu }  = n^{ \mu }\,,\\[5pt]
\label{eq:kinematics-with-pperp}
p_{ M }^{ \mu } &=  \alpha_{ M } n^{ \mu } + \frac{  |\vec{p}_{t}+ \frac{\vec{\Delta} _{t}}{2}|^2  +m_{ M }^{2}}{ \alpha _{M}s}p^{ \mu }-p_{\perp}^{ \mu }-\frac{ \Delta _{\perp}^{ \mu }}{2}\,,\quad\;\;
k^{ \mu } =  \alpha n^{ \mu } + \frac{  |\vec{p}_{t}- \frac{\vec{\Delta} _{t}}{2}|^2  }{ \alpha s}p^{ \mu }+p_{\perp}^{ \mu }-\frac{ \Delta _{\perp}^{ \mu }}{2}\,,
\end{align}
with $  \Delta  = p_{N'}-p_{N} $.
In the above, $ m_{N} $ is the nucleon mass, $ m_{M} $ is the meson mass, and $ \xi $ is the skewness parameter.
The centre-of-mass energy of the system, $ S_{\gamma N} $, is given by
\begin{align}
	S_{\gamma N} =  \left( q+p_{N} \right) ^2 =  \left( 1+\xi \right) s + m_{N}^2\,.
\end{align}
It is convenient to introduce the following Mandelstam variables:
\begin{align}
	t&=  \left( p_{N}-p_{N'} \right) ^2\,,\;\;\;\;u'=  \left( p_{M}-q \right) ^2\,,\;\;\;\;	t'= \left( k-q \right) ^2\,,\;\;\;\;	M_{\gamma M}^2 =  \left( p_{M}+k \right)^2\,.
\end{align}
In the Bjorken limit, where $  \vec{\Delta}_{t}  $, $ m_{N} $ and $ m_{M} $ are taken to zero, the kinematics simplifies to
\begin{align}
	\label{eq:Bjorken-kinematics}
	t = 0\,,\,
	 t' =  -\bar{ \alpha }M_{\gamma M}^{2}\,,\,
	  \alpha_{ M} =\bar{ \alpha }\equiv 1- \alpha \,,\,
	    \alpha  = \frac{-u'}{M^{2}_{\gamma M}}\,,\;\;\;
	    | \vec{p}_{t}|^2 =  \alpha \bar{ \alpha } M^{2}_{\gamma M}\,,\;\;\;
	      \xi = \frac{M_{\gamma M}^{2}}{2  S_{\gamma N}-M^{2}_{\gamma M} }\,,
\end{align}
We chose $ u' $, $ M_{\gamma M}^2 $ and $ S_{\gamma N} $ to be independent variables. The process factorises collinearly
when $  \vec{p}_{t}  $ is large, which in practice is ensured~\cite{Duplancic:2022ffo} by the following cuts on the Mandelstam variables
\begin{align}
	-u',\,-t'>1\,  \mathrm{GeV}^2\,,\qquad  -t < 0.5\,  \mathrm{GeV}^{2}\,, 
\end{align}
in order to avoid resonances between the outgoing meson and nucleon.

\subsection{Constructing the amplitude}

\label{sec:GPD-DA-models}

Our predictions rely on models for GPDs and DAs. GPDs are parametrised in terms of double distributions \cite{Radyushkin:1998es}. For the polarised (transversity) PDFs, which are used to construct the  $  \tilde{H} _{q} $ ($ H_{q}^{T} $) GPDs, we rely on the
``standard'' scenario, with flavour-symmetric light sea quark and antiquark distributions, and on the ``valence'' scenario, with densities taken to be completely flavour anti-symmetric.
For the DAs, we use the \textit{asymptotic} form and \textit{holographic} form\footnote{Suggested by AdS-QCD correspondence \cite{Brodsky:2006uqa}, dynamical chiral symmetry breaking on the light-front \cite{Shi:2015esa}, and recent lattice results \cite{Gao:2022vyh}.} 
\begin{align}
	\label{eq:DA-models}
	 \phi_{ \mathrm{asy} } (z) = 6 z  \left( 1-z \right) \,,\qquad \phi_{ \mathrm{hol} } (z) = \frac{8}{ \pi } \sqrt{z  \left( 1-z \right) }\,.
\end{align}
See~\cite{Duplancic:2022ffo,Duplancic:2023kwe} for the details on parameterisations.

In collinear factorisation, the amplitude $  {\cal A }  $ is computed through the convolution of the coefficient function $ T(x,z) $, the GPD $ H(x, \xi ,t) $, and the DA $  \phi (z) $,
\begin{align}
 {\cal A } = \int_{-1}^{1}dx \int _{0}^{1}dz\,T(x,z,\xi)\,H(x, \xi ,t)\, \phi (z)\,.
\end{align}
For each DA model \eqref{eq:DA-models}, we performed the $z$-integral  \textit{analytically}, and then the  $ x $-integral \textit{numerically}. Due to the rather small values of $\xi$, prefactors make dominant,  
for the chiral-even amplitude, the contributions from the GPDs $ H_{q} $ and $  \tilde{H}_{q}  $, and for the chiral-odd amplitude,  the contributions from the GPDs $ H_{q}^{T} $. See \cite{Duplancic:2022ffo} for the charged pion case, and \cite{Duplancic:2023kwe} for the  $ \rho  $ meson case.

\subsubsection{Unpolarised cross sections and counting rates}

After summing over the helicities of the incoming and outgoing particles, the squared amplitude is denoted by $ |\overline{ {\cal A } }|^2 $. The unpolarised differential cross section reads
\begin{align}
	\frac{d \sigma }{dt\, du'\, dM_{\gamma M}^{2}}\Bigg{|}_{(-t)=(-t)_{ \mathrm{min} }} = \frac{|\overline{ {\cal A } }|^2}{32 S_{\gamma N}^{2} M_{\gamma M}^{2} \left( 2 \pi  \right)^3 }\,,
\end{align}
where $ (-t)_{ \mathrm{min} }  = \frac{4 \xi ^2 m_{N}^2}{1- \xi ^2}$. We refer the reader to the original papers \cite{Duplancic:2022ffo,Duplancic:2023kwe} for plots corresponding to various  kinds of mesons and centre-of-mass energies ranging from JLab up to LHC and EIC.

Based on these results,
we performed a full phase space integration, taking into account the photon flux, so as to obtain the estimated number of events for the different mesons we consider, at various experiments. This is done for JLab (taking the integrated luminosity $ \int  {\cal L }dt  = 864 \, \mathrm{fb}^{-1}   $), future EIC (taking  $ \int  {\cal L }dt  = 10\; \mathrm{fb}^{-1}  $) and LHC in ultraperipheral collisions (UPCs) (assuming $ \int  {\cal L }dt  = 1200   \mathrm{nb}^{-1}  $). The obtained numbers are shown in Table \ref{tab:number-events}, for a proton target. The high energies available in a collider environment allows to perform a \textit{small $  \xi  $} study of quark GPDs, by restricting $ 300 < \frac{S_{\gamma N}}{\mathrm{GeV}^{2}} < 20000 $, which roughly
translates to $ 5 \cdot 10^{-5} <  \xi  < 5 \cdot 10^{-3} $. The expected number of events by employing this cut is also found in Table \ref{tab:number-events}.

\begin{table}
	\centering
\begin{tabular}{|c|c|c|c|}
\hline
Experiment & Meson & Number of events & Number of events with $ S_{\gamma N}>300  \mathrm{GeV}^2  $ \\
\hline
\hline
\multirow{3}{*}{JLab}
& $  \rho ^{0}_{L} $ & 1.3-2.4 $ \times 10^{5}$ & - \\\cline{2-4}
&$  \rho ^{0}_{T} $ & 2.1-4.2 $ \times 10^{4} $ & - \\\cline{2-4}
&$  \pi ^{+} $ &  0.3-1.8 $\times 10^{5}$ & - \\\hline
\multirow{3}{*}{EIC}
& $  \rho ^{0}_{L} $ & 1.3-2.4 $ \times 10^{4} $ & 0.6-1.2 $ \times 10^{3} $ \\\cline{2-4}
&$  \rho ^{0}_{T} $ & 1.2-2.4 $ \times 10^{3} $ & - \\\cline{2-4}
&$  \pi ^{+} $ & 0.2-1.3 $ \times 10^{4} $ & 1.4-5.0 $ \times 10^{2} $ \\\hline
\multirow{3}{*}{LHC in UPCs}
& $  \rho ^{0}_{L} $ & 0.9-1.6 $ \times 10^{4} $ & 4.1-8.1 $ \times 10^{2} $ \\\cline{2-4}
&$  \rho ^{0}_{T} $ & 0.8-1.7 $ \times 10^{3} $ & - \\\cline{2-4}
&$  \pi ^{+} $ & 1.6-9.3 $ \times 10^{3} $ & 1.0-3.4 $ \times 10^{2} $ \\\hline
\end{tabular}
\caption{The expected number of events for the photoproduction of a photon-meson at JLab, future EIC and LHC in UPCs is shown for the $  \rho ^{0}_{L} $, $  \rho ^{0}_{T} $ and $  \pi ^{+} $ cases, on a proton target. The number of events at small skewness $  \xi  $ ($ S_{\gamma N}> 300  \mathrm{GeV}^{2}  $) is also shown for EIC and LHC kinematics.}
\label{tab:number-events}
\end{table}

Thus, one finds that the statistics are very good, which warrants a proper experimental analysis of the process, especially at JLab, where the number of events can be as high as $ 10^5 $. For the chiral-odd case ($  \rho _{T} $), the cross section is proportional to $  \xi ^2 $, so that the expected number of events becomes negligible at small $  \xi  $, and hence they are omitted from the table.

\subsubsection{Polarisation asymmetries with respect to the incoming photon}

\begin{figure}
	\centering
	\psfrag{HHH}{\hspace{-1.5cm}\raisebox{-.5cm}{\scalebox{.8}{$ M^{2}_{\gamma  \rho ^{+}} ({\rm 
				GeV}^{2})$}}}
\psfrag{VVV}{ $ \mathrm{LPA}_{{  \gamma { \rho }^{+}_{L} }} $ }
\psfrag{TTT}{}
	\includegraphics[width=0.46\textwidth]{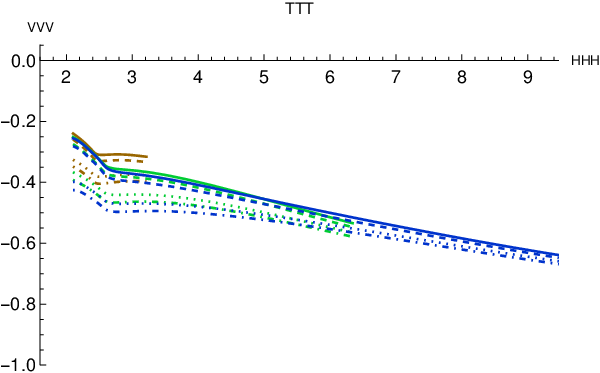}
	\hfill
	\psfrag{HHH}{\hspace{-1.5cm}\raisebox{-.5cm}{\scalebox{.8}{$M^{2}_{\gamma  \pi ^{+}} ({\rm 
					GeV}^{2})$}}}
	\psfrag{VVV}{$ \mathrm{LPA}_{{ \gamma { \pi ^{+}} }} $}
	\psfrag{TTT}{}
	\includegraphics[width=0.46\textwidth]{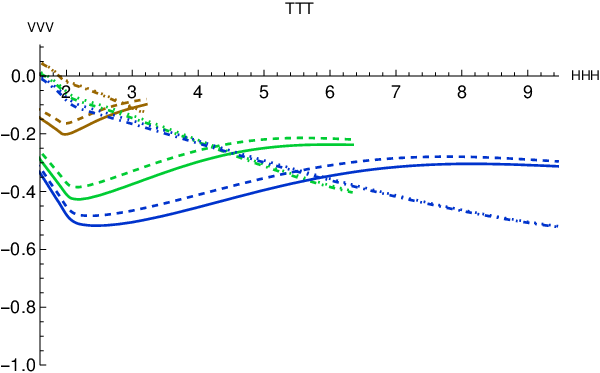}
	\caption{The LPA, constructed from differential cross sections wrt $ M_{\gamma M}^{2} $, is shown as a function of $ M_{\gamma M}^{2} $ for the $   \rho ^{+}  $ (left) and $  \pi ^{+} $ cases. In both plots, dashed (non-dashed) corresponds to holographic (asymptotical) DA, while dotted (non-dotted) corresponds to standard (valence) scenario for the GPD model used. }
	\label{fig:LPA}
\end{figure}

Besides the cross sections, we also computed polarisation asymmetries wrt the incoming photon. Due to the invariance of QCD/QED under parity~\cite{Duplancic:2022ffo}, the \textit{circular} polarisation asymmetry vanishes for all the mesons that we considered. Instead, we calculated the \textit{linear} polarisation asymmetry (LPA), using the \textit{Kleiss-Stirling spinor techniques}. This linear polarisation asymmetry also vanishes for the case of a transversely polarised $  \rho  $-meson. The LPA reads
\begin{align}
	\label{eq:LPA-def}
	 \mathrm{LPA} = \frac{d \sigma _{x}-d \sigma _{y}}{d \sigma _{x}+d \sigma _{y}}\,,
\end{align}
where $ d \sigma  $ is a differential or integrated cross section, and the subscript $ x $ (or $ y $) represents the direction of polarisation of the incoming photon, in a frame where the $ x $-direction is defined by the direction of the outgoing photon.
In practice, it is more natural to deal with the measured asymmetry in the lab frame, denoted by $  \mathrm{LPA}_{ \mathrm{Lab} }  $. The latter is related to the above LPA in \eqref{eq:LPA-def} by
\begin{align}
\mathrm{LPA}_{ \mathrm{Lab} } = \mathrm{LPA}\cos  \left( 2 \theta  \right) \,,
\end{align}
where $  \theta  $ is the angle between the lab frame $ x $-direction and $ p_{\perp} $, which varies event by event.
In Fig.~\ref{fig:LPA} we show the dependence of the LPA, constructed from the differential cross section in $ M_{\gamma M}^{2} $, as a function of $ M_{\gamma M}^2 $ for the $  \rho ^{+}_{L} $ meson (left) and $  \pi ^{+} $ meson (right) cases. The LPA is sizeable in both cases, reaching up to $ 60\% $.
For the $  \pi ^{+} $ case, the LPA could be used for disentangling GPD models, while the effect coming from the DA models is suppressed. 

\section{Conclusions}

The exclusive photoproduction of a photon-meson pair is a very promising channel to study quark GPDs.
This is not only an additional channel for probing quark GPDs, but also one of the few processes for extracting chiral-odd GPDs at the leading twist, by choosing the outgoing meson to be a transversely polarised $  \rho  $-meson. Our estimation of the statistics shows that measurements at JLab, COMPASS, future EIC and LHC in ultraperipheral collisions are very promising. In particular, we estimate for JLab a number of events of the order of $ 10^5 $. We also obtain sizeable LPA.

\providecommand{\href}[2]{#2}\begingroup\raggedright\endgroup

\end{document}